# Thermal Escape from a Metastable State in Periodically Driven Josephson Junctions

Guozhu Sun, Jian Chen, Weiwei Xu, Zhengming Ji, Lin Kang, Peiheng Wu,

Research Institute of Superconductor Electronics, Department of Electronic Science and Engineering, Nanjing University, Nanjing, 210093, China,

Guangfeng Mao, Ning Dong, Yang Yu[*], and Dingyu Xing

National Laboratory of Solid State Microstructures and Department of Physics, Nanjing University, Nanjing, 210093, China.

## Abstract

Resonant activation and noise-enhanced stability were observed in an underdamped real physical system, i.e., Josephson tunnel junctions. With a weak sinusoidal driving force applied, the thermal activated escape from a potential well underwent resonance-like behavior as a function of the driving frequency. The resonance also crucially depended on the initial condition of the system. Numerical simulations showed good agreement with the experimental results.

The thermal activated escape from a metastable state is a ubiquitous problem in physics, chemistry, and biology [1,2]. The escape process can be well understood within Kramers theory. Recently, the investigation of nonlinear system in the presence of noise showed some counterintuitive resonance-like phenomena. Typical examples are stochastic resonance (SR) [3], resonant activation (RA) [4,5], and noise-enhanced stability (NES) [6,7,8]. The signature of all these phenomena is a nonmonotonic behavior of some quantity, e.g. the lifetime of a metastable state, as a function of the noise intensity or the driving frequency. However, the previous studies were focused on the overdamped systems. Particularly, the experimental investigation in the underdamped physical system was sparse so far.

In this letter, we experimentally observed the coexistence of RA and NES in underdamped Josephson tunnel junctions. The escape rates could be enhanced or suppressed by a weak periodic force with different initial conditions. Although Mantegna *et al.* have observed NES in an overdamped unstable system using a circuit model [8], this is the first report of NES in underdamped metastable systems to the best of our knowledge. Our investigation indicated that the mechanisms of RA and NES are the same. Moreover, we experimentally determined the initial conditions of RA and NES, which agreed with that of the numerical simulations remarkably well.

The equation of motion of a current-biased Josephson tunnel junction is identical to the classical equation of motion of a particle moving in a washboard potential [9]

$$\frac{\hbar C}{2e}\frac{d^2\delta}{dt^2}+\frac{\hbar}{2eR}\frac{d\delta}{dt}+I_c\sin\delta=I_b+I_n(t) \tag{1}$$

where $\delta$ is the phase difference across the junction, $C$ is the capacitance of the junction, $R$ is the model shunting resistance, $I_b$ is the bias current, and $I_c$ is the critical current. The term $I_n(t)$ represents the thermal noise current due to $R$, which is related with $R$ and the temperature $T$ as

$$\langle I_n(t) I_n(t') \rangle = 2\frac{k_B T}{R}\delta(t-t') \tag{2}$$

For $I_b < I_c$, the potential has a series of metastable wells with barrier height

$$\Delta U = 2E_J(\sqrt{1-i^2} - i\cos^{-1} i) \tag{3}$$

where $i = I_b/I_c$ is the normalized bias current, $E_J = I_c \Phi_0/2\pi$ is the Josephson coupling energy, and $\Phi_0 = h/2e$ is the flux quantum. A junction initially trapped in a potential well (corresponding a zero voltage drop on the junction) can be activated out of the well by thermal fluctuation. The lifetime of the zero voltage state is given by [10]

$$\tau \equiv 1/\Gamma_t = \left(a_t \frac{\omega_p}{2\pi}\right)^{-1} \exp(\Delta U/k_B T), \tag{4}$$

with $\omega_p = \omega_{p0}(1-i^2)^{1/4} \equiv \sqrt{2\pi I_c/\Phi_0 C}(1-i^2)^{1/4}$ being the small oscillation frequency of the particle at the bottom of the well, and $k_B$ being the Boltzmann's constant. The prefactor $a_t$ is weakly dependent on the damping of the system and can be treated as a constant $a_t \approx 1$ in our case [1,11].

The samples used in this study were NbN/AlN/NbN Josephson tunnel junctions. The critical temperature $T_c$ was about 16 K. The $I-V$ characteristic measurements showed that the junctions were of high quality with large hysteresis, indicating the system was a highly underdamped system [12]. The sizes of the junction were 6 μm × 6 μm. The critical current and the capacitance of the junction were $I_c \sim$ 1.6 mA and $C \sim$ 1.8 pF. The sample was mounted on a chip carrier that was enclosed in a helium-filled oxygen-free copper sample cell, which was then dipped into liquid helium. The junction was magnetically shielded by a Mu-metal cylinder. All electrical leads that connected the junctions to room temperature electronics were carefully filtered by RC filters (with a cutoff frequency of about 1 MHz) and Cu-powder filters (with a cutoff frequency of about 1 GHz). Battery-powered low-noise preamplifiers were used for all measurements. The experiments were performed in a copper shielded room.

The escape times of the junction were directly measured using the time-domain technique [13]. For each escape event, we started the cycle by ramping up the bias current to a value $I_b$, which was very close to $I_c$ ($I_b \sim 0.99 I_c$), and maintaining at this level for a period of waiting time. A sinusoidal current $I_{ac} = 2i_{ac}\sin(2\pi f t + \varphi_0)$ was added to the bias current during the waiting time, where $t=0$ when the bias current reached $I_b$. Eq. 1 now reads

$$\frac{\hbar C}{2e}\frac{d^2\delta}{dt^2} + \frac{\hbar}{2eR}\frac{d\delta}{dt} + I_c \sin\delta = I_b + I_{ac} + I_n(t) \qquad (5)$$

In our experiments, $i_{ac} \ll I_b$ and $I_b + 2i_{ac} < I_c$, the junction was still a metastable system. In addition, $2\pi f \ll \omega_p$, prohibiting the dynamic resonant escape from the system [14]. The junction voltage was fed to a timer, which was triggered by the sudden voltage jump when the junction switched from zero-voltage state to finite-voltage state, to record escape time $t_{esc}$. The bias current $I_b$ was then decreased to zero, returning the junction to the zero-voltage state （Fig.1b）. The process was repeated about $10^4$ times to obtain an ensemble of the escape time. The average time of the ensemble represented the mean thermal activation escape time $<t_{esc}>$. Then we changed $f$, and measured another $<t_{esc}>$. By plotting $<t_{esc}>$ as a function of $f$ we observed a nonmonotonic behavior, shown in Fig. 2a. The underlying physics was that the particle actually saw a periodically oscillating barrier (Eq. 5). The strong correlation between the thermal activation and the barrier oscillation resulted the resonant escape (a minimum in $<t_{esc}>$), namely RA. The resonant frequencies $f_{res}$ corresponds to the escape time of the particle from the low configuration of the barrier. Therefore, when we increased $I_b$, the escape rate increased, leading to an increase of $f_{res}$ (Fig. 2b). From the best fit in Fig. 2b, we obtained that the matching condition was

$$4<t_{esc}> \simeq 1/f_{res}, \qquad (6)$$

which was similar to that reported in Ref. [5], where RA has been observed in an underdamped metastable system with a piece wise oscillating barrier. The presence of RA in both cases agrees with the theoretical analysis in overdamped systems [4], suggesting the availability of RA in all fluctuating barrier systems.

An interesting phenomenon here was that RA depends on the initial phase $\varphi_0$. When we changed the initial phase gradually, the minimum vanished. Moreover, for $\varphi_0 = \pi$, resonant peaks present at $f_{res}$ (Fig. 3a). The junction actually had a longer average escape time than that without periodical driving force. Therefore, the noise enhanced the stability of the metastable state. This phenomenon, named NES, was found in overdamped nonlinear systems recently [6,7,8]. Similar to that of RA, the origin of NES is the strong correlation between the thermal escape and the barrier fluctuation. We supported this argument by investigating the linear relation between $f_{res}$ and the bias current (i.e., the escape rate of the particle), shown in Fig. 3b. For both RA and NES, $f_{res}$ was proportional to <$t_{esc}$> with an almost identical prefactor. In addition, it was predicted that NES depends sensitively on the initial position of the particle [7]. In our experiment, the initial position is almost fixed because $2i_{ac} \ll I_b$. However, the initial phase is crucial to generate NES. We measured the average lifetime at various initial phases and found that the system exhibited RA behavior for $-\pi/3 < \varphi_0 < \pi/3$, but NES behavior for $5\pi/6 < \varphi_0 < 9\pi/6$.

While it is quite challenging to analytically solve Eq. 5 for the underdamped system, we performed numerical simulations in order to confirm our experimental observations. Shown in Fig. 4 and Fig. 5 are some examples of the simulation results where we can see clearly all the features of RA and NES presented in the experiments. In addition, the initial conditions for RA and NES obtained from simulations were very close to the experimental data.

Our experimental and numerical evidence indicated that RA and NES may happen in all dynamic systems, including both overdamped and underdamped systems. Currently, Josephson junctions are widely used as detector in the rapid single flux quantum (RSFQ) circuitry [15] and superconducting quantum bits (qubits) [16]. Our study is not only important to understand the physics of fluctuations in a Josephson junction to improve the performance of the devices, but also in nonequilibrium statistical mechanics of low dissipative systems. Moreover, underdamped systems [16] are employed to build superconducting qubits due to the requirement of the long decoherence time. It is very interesting that NES provides a quite unexpected way to maximize the stability of qubits.

In summary, we experimentally observed the coexistence of RA and NES in underdamped Josephson tunnel junctions. It was found that RA and NES are generated by the correlation between the thermal activation and the barrier fluctuation. The different initial conditions leaded to RA and NES respectively. The numerical simulations confirmed our observation, indicating that Josephson junctions are ideal vehicle for probing relevant physics issues in metastable systems.

This work is partially support by the National Natural Science Foundation of China（10474036）, the National Basic Research Program of China, 973 Program （2006CB61006） and the Doctoral Funds of Ministry of Education of the People's Republic of China (2004028033).

# Figure Captions

Fig. 1. (a) Washboard potential and equivalent circuit (inset) of a current biased Josephson tunnel junctions with $I_b < I_c$. (b) Schematic time profile for measuring the escape time of a Josephson junction subjected to a weak sinusoidal force.

Fig. 2. (a) Average escape time as a function of the driving frequency at 4.2 K for various normalized bias currents which are marked below the curves. The initial phase was $\varphi_0 = 0$, and $i_{ac} \sim 0.002\ I_b$. The valleys indicate RA with the resonant frequency depending on the bias current. (b) The minimum average escape time vs. inverted resonant frequency. The dashed line is the best fit with a slope of $0.26 \pm 0.03$.

Fig. 3. (a) Average escape time as a function of the driving frequency for different bias currents (marked on the curves) with the initial phase $\varphi_0 = \pi$. The temperature was 4.2 K and $i_{ac} \sim 0.002\ I_b$. Resonant peaks, instead of valleys, were clearly observed, indicating NES in the system. (b) The maximum average escape time vs. inverted resonant frequency. The dashed line is the best fit with slope $0.23 \pm 0.01$.

Fig. 4. The numerically calculated average escape time vs. the frequency of the sinusoidal driving force with $\varphi_0 = 0$. The parameters used in the simulations were: $T$ = 4.2 K, $C$ = 1.8 pF, $R$ = 2 ohm, $I_c$ = 1.6 mA, $i_{ac} \sim 0.002\ I_c$. The minimum of the average escape time moved to the higher frequency when the bias current increased, consistent with the experimental observations. The matching condition obtained from the calculation was $<t_{esc}> \simeq 0.23/\ f_{res}$, showing quantitative agreement with the experiment observations.

Fig. 5. The numerically calculated average escape time vs. the frequency of the sinusoidal driving force with $\varphi_0 = \pi$. The maximum of NES moved to higher frequency as the increasing of the bias current. The parameters used were the same as those used in Fig. 4.

# References

*yuyang@nju.edu.cn

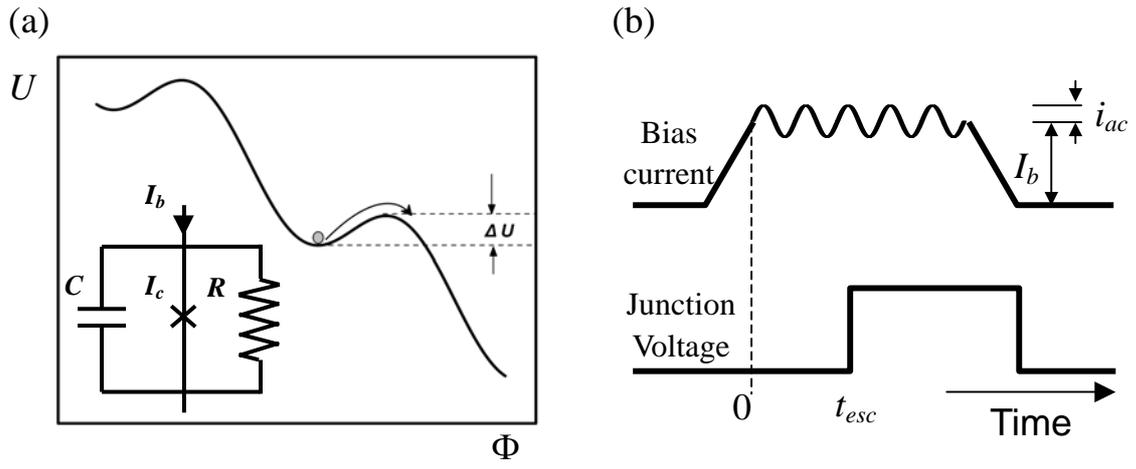

Fig. 1   G. Sun *et al.*

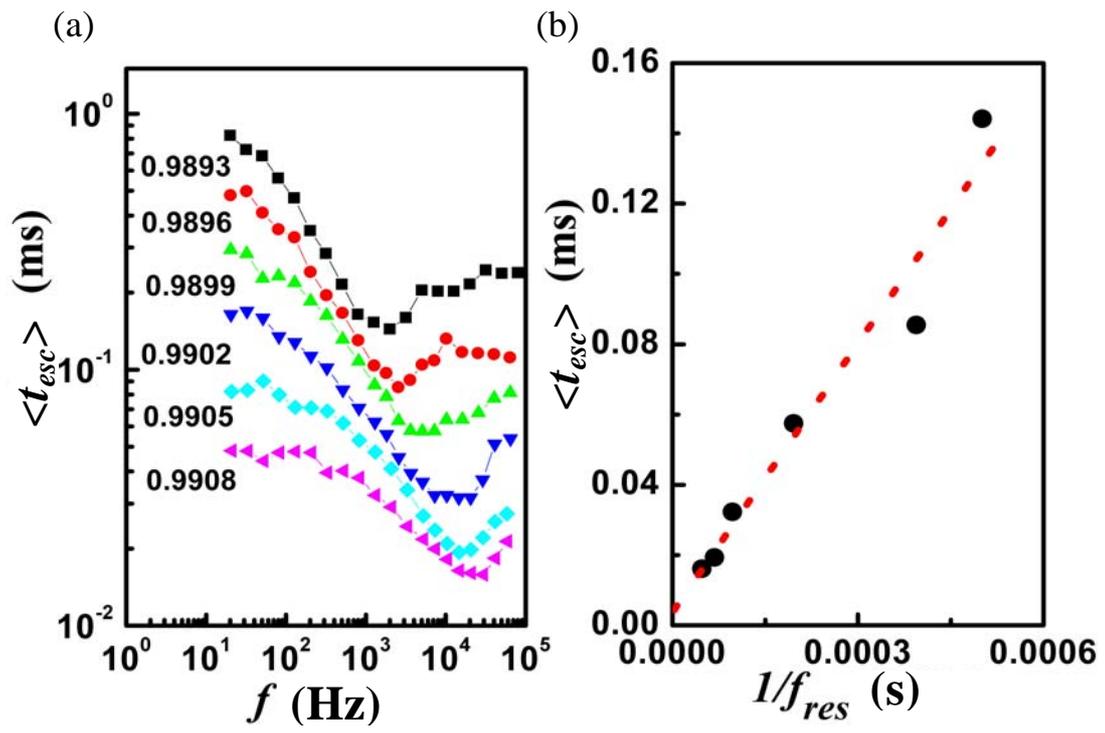

Fig. 2 G. Sun *et al.*

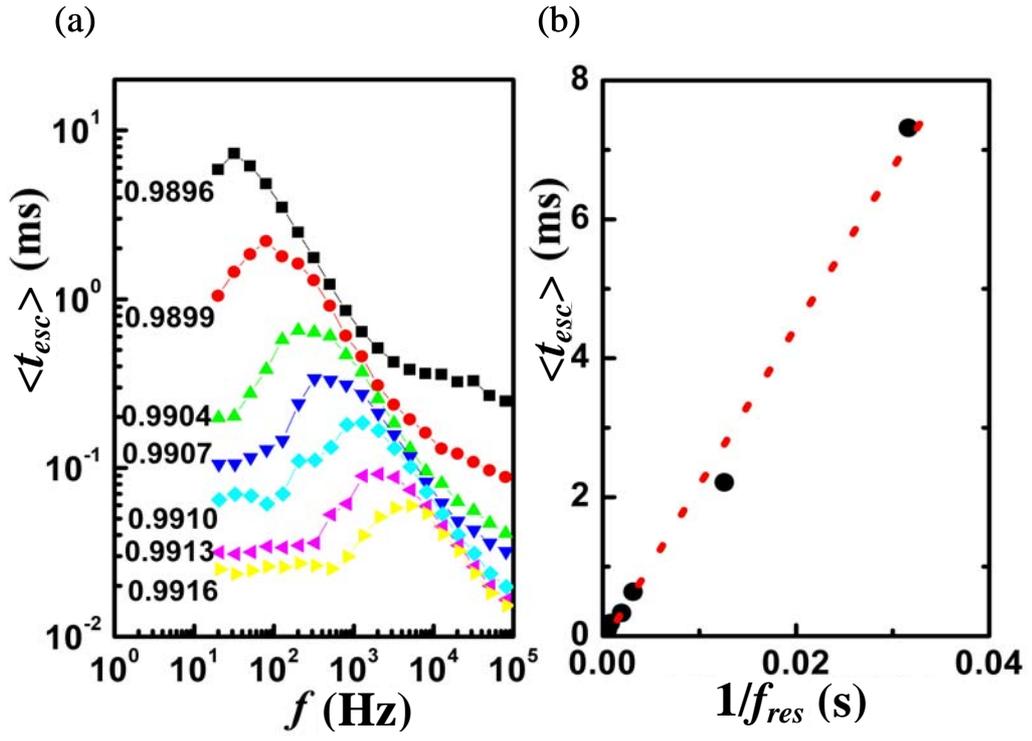

Fig. 3  G. Sun *et al.*

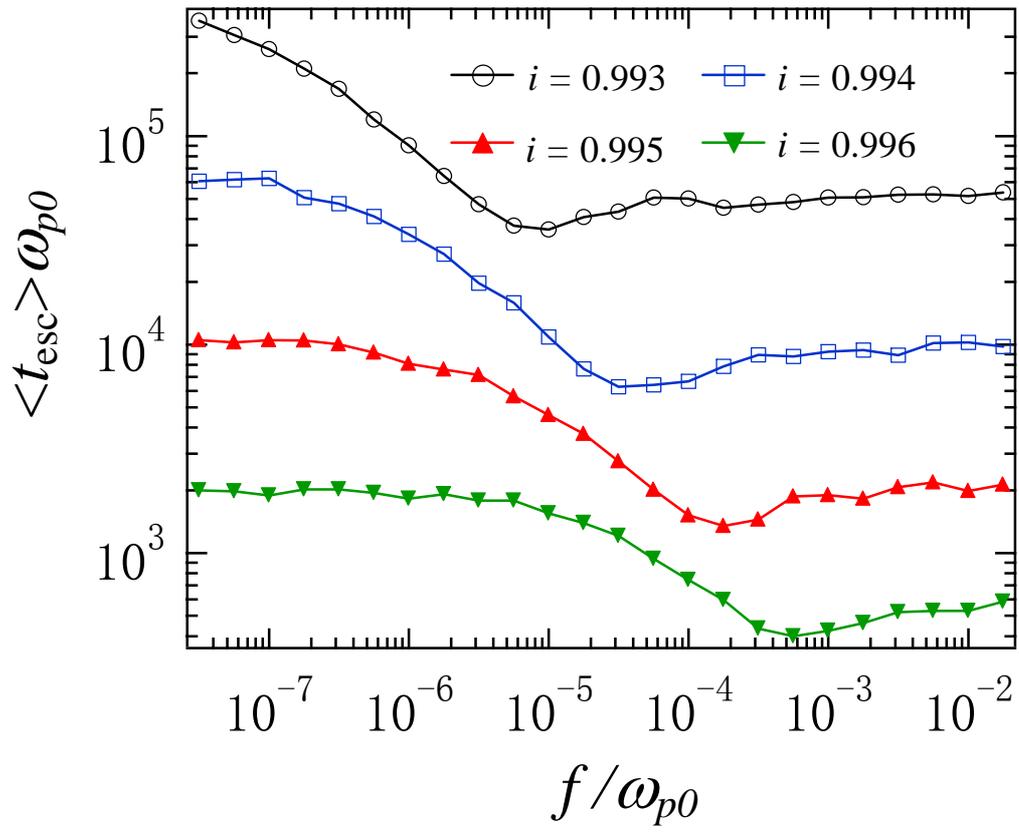

Fig. 4  G. Sun *et al.*

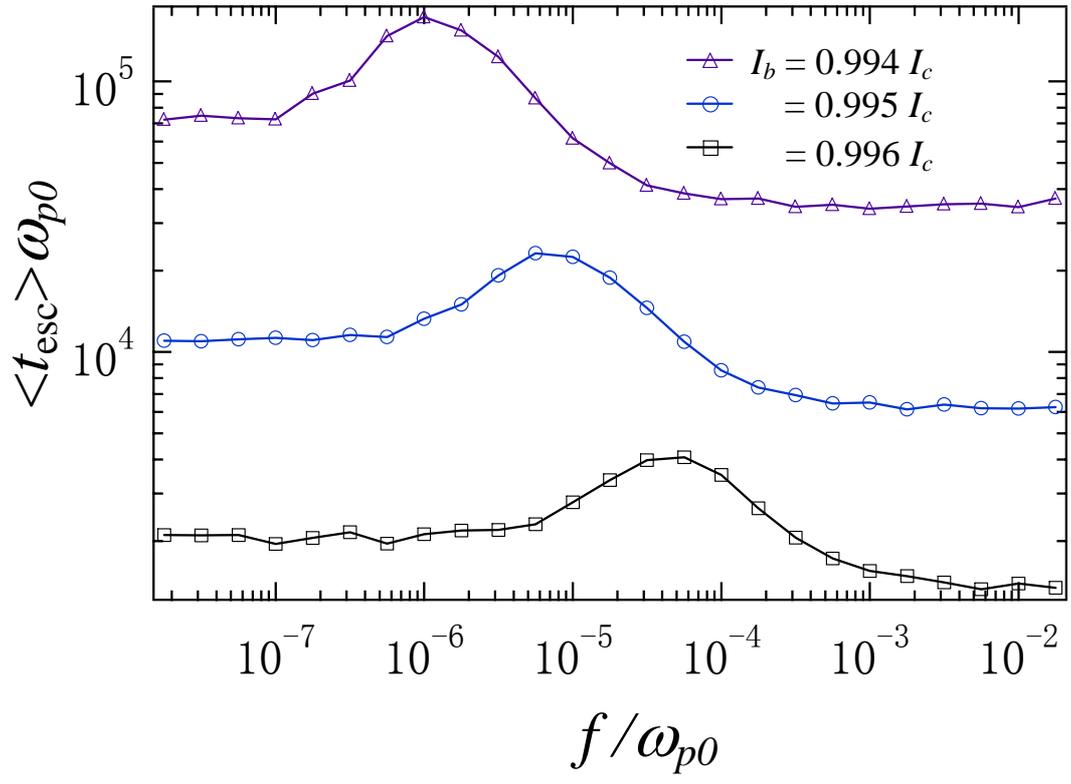

Fig. 5　G. Sun *et al.*